\documentclass[11pt,a4paper,epsfig]{article}
\usepackage{amsmath}
\usepackage{amsfonts}
\usepackage{epsfig}
\begin{document}

\title{Elementary solutions of the quantum planar\\ two center problem}

\author{M. A. Gonz\'alez Le\'on$^1$, J. Mateos Guilarte$^2$ and M. de la Torre Mayado$^2$ \\ {\sl \small
$^1$ Departamento de Matem\'atica Aplicada, Universidad de Salamanca, Spain}\\
{\sl \small $^2$ Departamento de F\'{\i}sica Fundamental, Universidad de Salamanca, Spain}}

\date{}

\maketitle

\abstract{The quantum problem of an electron moving in a plane under the field created by two Coulombian centers admits simple analytical solutions for some particular inter-center distances. These elementary eigenfunctions, akin to those found by Demkov for the analogous three dimensional problem, are calculated using the framework of quasi-exact solvability of a pair of entangled ODE's descendants from the Heun equation. A different but interesting situation arises when the two centers have the same strength. In this case completely elementary solutions do not exist.}

\section{Introduction}

The quantum spectral problem of a charged particle moving in the field created by two fixed Coulomb centers appears in a natural way in the study of diatomic molecular ions, in the Born-Oppenheimer approximation. Although the three-dimensional case has been profusely studied from different points of view: computational, see e.g. \cite{Ponomarev}, analytical, see \cite{Leaver2, Liu},  as far as we know the search for elementary solutions has been reduced to the work of Demkov \cite{Demkov} in the two-center problem in three dimensions.

In this letter, the existence of elementary solutions for the planar problem is studied. The underlying idea is similar to the Demkov approach \cite{Demkov} to the three-dimensional case, and thus we search for the eigenfunctions of the system that can be written essentially as polynomials. Unlike the three-dimensional problem, the planar system admits elementary solutions not only of hydrogenoid type, i.e., eigenfunctions with energy coinciding with the energy levels of an hydrogenoid atom, but also of a new type that we shall term as \lq\lq quasi\rq\rq-hydrogenoid. These new type of elementary solutions does not arise in the three dimensional problem, its existence in the plane is reminiscent of the particular topology subjacent to the 2D problem.

The Schr\"odinger equation for the planar problem is separable in Euler elliptic coordinates. After the separation process the PDE wave equation reduces to two ODE's, namely a Razavy and a Whittaker-Hill equation, linked by the two invariants of the system. In his original article \cite{Razavy} Razavy showed that, for certain values of parameters, a finite part of the spectrum was calculable in closed form. Later, in \cite{Khare, Konwent}, it was proved that these exact computable eigenfunctions for Razavy eq. could be obtained in terms of four polynomial sets. Finkel et al \cite{Finkel1999} showed that the Schr\"odinger equation associated to the Razavy potential is quasi-exactly solvable (QES), the mentioned sets of polynomials was nothing but instances of the associated weakly orthogonal polynomial family \cite{Finkel1996}, which allow to find these special eigenfunctions. Finally, these results were also extended to the Whittaker-Hill eq. in \cite{Finkel1999}.

In this work we will follow a slightly different approach treating the two equations as particular descendants of the Confluent Heun equation (CHEq), and, thus, considering that it is possible to obtain polynomial solutions when the parameters of the problem render these two versions of CHEq simultaneously QES \cite{CHEqnosotros}.

The separability properties and the study of the general potentials that admit exact and quasi-exact solvability for a broad family of physical problems associated to the 3D two-center problem have been recently analyzed thoroughly in ref. \cite{Turbiner2}.

When the two centers are of equal strength, contrary to expectations, the problem is more involved. The Whittaker-Hill equation descends to the Mathieu equation which is never QES. Thus, there are not completely elementary eigenfunctions in this situation, although interesting wave functions may be derived for certain intercenter distances determined from the Mathieu characteristic values.

\section{The planar quantum problem posed by two Coulombian centers }

The stationary Schr$\ddot{\rm o}$dinger equation governing the quantum dynamics of a charged particle moving in the potential of two fixed Coulombian centers (nuclei, ions, etc) in the plane reads:

\begin{equation}
\label{Sch2c2D}
\left( -\frac{1}{2} \Delta  -  \frac{Z_1}{r_1} - \frac{Z_2}{r_2}\right)  \Psi  = E \Psi  \, ,
\end{equation}
where $\Delta=\frac{\partial^2}{\partial x_1^2}+\frac{\partial^2}{\partial x_2^2}$, and atomic units are used: $\hbar = m_e = e = a_0 = 1$, $a_0$ being the Bohr radius. $Z_1$ and $Z_2$ are the atomic numbers of the two nuclei, and $r_1$ and $r_2$ are the distances from the electron to the nuclei:

\begin{equation}
r_1=\sqrt{\left(x_1-\frac{R}{2}\right)^2+x_2^2}\, ,\  r_2=\sqrt{\left(x_1+\frac{R}{2}\right)^2+x_2^2}\,  ,\nonumber
 \end{equation}
while $R$ is the internuclear distance. Equation (\ref{Sch2c2D}) admits separation of variables using Euler elliptic coordinates $(\xi,\eta)$, $\xi= (r_1+r_2)/R\in(1,+\infty)$ and $\eta = (r_2-r_1)/R \in(-1,1)$. Search for separated wave functions $\Psi(\xi,\eta) = F(\xi)  G(\eta)$ converts eq. (\ref{Sch2c2D}) into two ODEs:
one
\begin{eqnarray}
\label{radial2d}
&&(\xi^2-1)\frac{d^2F(\xi)}{d\xi^2}+ \xi \frac{dF(\xi)}{d\xi} \nonumber\\
&&+\left( \frac{ER^2}{2} \xi^2+R(Z_1+Z_2) \xi+\lambda\right) F(\xi) = 0
\end{eqnarray}
in the \lq\lq radial\rq\rq coordinate $\xi\in (1,\infty)$, the other
\begin{eqnarray}
\label{angular2d}
&&(1-\eta^2)\frac{d^2G(\eta)}{d\eta^2}- \eta \frac{dG(\eta)}{d\eta}\nonumber \\
&&- \left( \frac{ER^2}{2} \eta^2+R(-Z_1+Z_2) \eta+\lambda\right) G(\eta) = 0
\end{eqnarray}
in the \lq\lq angular\rq\rq $\eta\in (-1,1)$ coordinate. $\lambda$ is the separation constant.

Equations (\ref{radial2d}) and (\ref{angular2d}) are no more than the algebraic form of Razavy and Whittaker-Hill equations respectively, see e.g. \cite{GGTCM}. The wave functions of our problem are given by the product of solutions of (\ref{radial2d}) and (\ref{angular2d})
for identical values of the energy $E$ and the separation constant $\lambda$.

\subsection{The connection with the Confluent Heun equation}

Both (\ref{radial2d}) and (\ref{angular2d}) are reduced to CHEq \cite{Ronveaux,Slavyanov}:
\begin{eqnarray}
&&\left( z^2-1\right)u''(z)+ \left( \frac{\epsilon}{2} (z^2-1)+ \gamma (z-1)\right. \nonumber \\ && \left. +\delta(z+1)\right) u'(z)+ \left( \frac{\alpha}{2} (z+1)-q\right)u(z)  = 0 \, ,\label{CHEs}
\end{eqnarray}
via the changes of variable:

\begin{eqnarray}
F(\xi)&=& (\xi+1)^{\frac{2\gamma-1}{4}}\, (\xi-1)^{\frac{2\delta-1}{4}}\, e^{\frac{\epsilon \xi}{4}}\, u(\xi)\\
G(\eta)&=&(1+\eta)^{\frac{2\gamma-1}{4}}\, (1-\eta)^{\frac{2\delta-1}{4}}\, e^{\frac{\epsilon \eta}{4}}\, u(\eta)
\end{eqnarray}
that are compatible with the form of Razavy and Whittaker-Hill equations only in the four following cases: a) $\delta=\gamma=1/2$, b) $\delta=\gamma=3/2$, c) $\delta=1/2,\  \gamma=3/2$, and d) $\delta=3/2,\  \gamma=1/2$. The rest of CHEq constants are related to the physical parameters of eqs. (\ref{radial2d}) and (\ref{angular2d}) by the identities: $\epsilon^2= -8ER^2$, and

\begin{equation}
\lambda=\frac{\epsilon^2}{16} + \frac{\epsilon(\gamma-\delta)}{4} -\frac{(\gamma+\delta)(\gamma+\delta-2)}{4} + \frac{2\alpha-1}{4}-q\nonumber
\end{equation}
together with:

\begin{equation}
\frac{\alpha}{2}-\frac{\epsilon}{4}(\delta+\gamma)=R(Z_1+Z_2)\nonumber
\end{equation}
in the radial equation, and

\begin{equation}
\frac{\alpha}{2}-\frac{\epsilon}{4}(\delta+\gamma)=R(Z_2-Z_1)\nonumber
\end{equation}
in the angular one.

If we interpret CHEq as an spectral problem $Du(z)=qu(z)$, it is not difficult to show, see \cite{CHEqnosotros}, that the differential operator $D$ can be written as a quadratic combination of the generators of the Lie algebra $sl(2,R)$ if and only if $\alpha=-n\epsilon$, being $n$ a non-negative integer, and consequently CHEq represents a quasi-exactly solvable spectral problem \cite{Turbiner, Ushveridze, TurShifman} for this special combination of parameters. In this situation it is possible to find an invariant module of polynomial solutions of (\ref{CHEs}) associated with each arbitrary value of $n$. Following the standard procedure \cite{Finkel1996, Bender} we search for Frobenius solutions of the form:

\begin{equation}
u(z) =   \sum_{k=0}^\infty \frac{(-1)^k   {\cal P}_k(q)}{2^{k} k! (\gamma)_k }  (z+1)^k \label{frob1}\, ,
\end{equation}
where $(\gamma)_k=\gamma (\gamma+1)\dots (\gamma+k-1)$ and ${\cal P}_0(q)=1$, which leads to the following three-term recurrence between the polynomials ${\cal P}_k(q)$ for $k\geq 1$:
\begin{eqnarray}
{\cal P}_{k+1}(q) &=& \left( q-k(\delta+\gamma-\epsilon+k-1)\right)  {\cal P}_k(q)\nonumber\\ && - k \epsilon  (n-k+1)(\gamma+k-1)  {\cal P}_{k-1}(q)\label{recurrence}
\end{eqnarray}
Thus, given a concrete value of $n$, and fixing $q$ as one of the $n+1$ roots of ${\cal P}_{n+1}(q)$, $q_j$, $j=1,\dots,n+1$, solutions (\ref{frob1}) truncate to polynomials of degree $n$,

\begin{equation}
u_{n,j}(z)=\sum_{k=0}^n \frac{(-1)^k   {\cal P}_k(q_j)}{2^{k} k! (\gamma)_k }  (z+1)^k
\end{equation}
and consequently a $n+1$ dimensional module of polynomial eigenfunctions of (\ref{CHEs}) can be determined algebraically. The polynomials ${\cal P}_{k}(q)$ constitute the weakly orthogonal polynomial family \cite{Finkel1996} associated to the QES property for CHEq \cite{CHEqnosotros}. It is interesting to remark that the particularization of this family to the Razavy and Whittaker-Hill parameters corresponds with the first polynomial family found in \cite{Finkel1999} for these equations.

\section{Elementary solutions of Razavy equation}

Translating the QES condition of (\ref{CHEs}) to the Razavy equation (\ref{radial2d}): $\alpha= - n^r \epsilon$, $n^r \in {\mathbb N}$, a quantization condition in the expression of $\epsilon$ appears

\begin{equation}
\frac{\alpha}{2}-\frac{\epsilon}{4}(\gamma+\delta) = R(Z_1+Z_2) \Longrightarrow  \epsilon = - \frac{4 R (Z_1+Z_2)}{(2 n^r + \gamma+\delta)}
\end{equation}
and consequently, in the energy eigenvalues:

\begin{equation}
E=-\frac{\epsilon^2}{8 R^2} \Longrightarrow  E_{n^r}=-\frac{2 (Z_1+Z_2)^2}{(2 n^r +\gamma+\delta)^2}\label{energyRaz}
\end{equation}
Meanwhile, the separation constant is characterized by $n^r$ but also by $j=1,\dots,n^r+1$, because the explicit dependence in the $q_j$ parameter:
\begin{eqnarray}
\lambda_{n^r,j} &=& \frac{R^2(Z_1+Z_2)^2}{(2 n^r +\gamma+\delta)^2} + \frac{(2 n^r -\gamma+\delta)R(Z_1+Z_2)}{(2 n^r +\gamma+\delta)} \nonumber\\ && -\frac{(\gamma+\delta)(\gamma+\delta-2)}{4} -\frac{1}{4}-q_j\label{lambdaj}
\end{eqnarray}
As it was mentioned above, the compatibility between Razavy and CHEq is only possible in four concrete choices of constants $\delta$ and $\gamma$. Thus we find the following expressions for energies, separation constants and eigenfunctions:

\textbf{Type a}. $\delta=\gamma=\frac{1}{2}$.
\begin{eqnarray}
 E_{n^r}&=&-\frac{2 (Z_1+Z_2)^2}{(2 n^r +1)^2}\nonumber\\ \lambda_{n^r,j}&=& -\frac{R^2E_{n^r}}{2} + \frac{2 n^r R(Z_1+Z_2)}{2 n^r +1} -q_j\nonumber\\ F_{n^r,j}(\xi)&=& e^{- \frac{R(Z_1+Z_2)}{2n^r+1}\xi} u_{n^r,j}(\xi)\nonumber
\end{eqnarray}

\textbf{Type b}. $\delta=\gamma=\frac{3}{2}$.
\begin{eqnarray}
 E_{n^r}&=&-\frac{2 (Z_1+Z_2)^2}{(2 n^r +3)^2}\nonumber\\ \lambda_{n^r,j}&=& -\frac{R^2E_{n^r}}{2} + \frac{2 n^r R(Z_1+Z_2)}{2 n^r +3} -q_j-1\nonumber\\ F_{n^r,j}(\xi)&=& \sqrt{\xi^2-1}\,  e^{- \frac{R(Z_1+Z_2)}{2n^r+3}\xi} u_{n^r,j}(\xi)\nonumber
\end{eqnarray}

In these two cases a) and b) the energies are the corresponding to planar hydrogenlike atoms with nuclear charge $Z_1+Z_2$.

\textbf{Type c}. $\delta=\frac{1}{2}$, $\gamma=\frac{3}{2}$.
\begin{eqnarray}
 E_{n^r}&=&-\frac{2 (Z_1+Z_2)^2}{(2 n^r +2)^2}\nonumber\\ \lambda_{n^r,j}&=& -\frac{R^2E_{n^r}}{2} + \frac{(2 n^r-1) R(Z_1+Z_2)}{2 n^r +2} -q_j-\frac{1}{4}\nonumber\\F_{n^r,j}(\xi)&=& \sqrt{\xi+1} \, e^{- \frac{R(Z_1+Z_2)}{2n^r+2}\xi} u_{n^r,j}(\xi)\nonumber
\end{eqnarray}

\textbf{Type d}. $\delta=\frac{3}{2}$, $\gamma=\frac{1}{2}$.
\begin{eqnarray}
 E_{n^r}&=&-\frac{2 (Z_1+Z_2)^2}{(2 n^r +2)^2}\nonumber\\ \lambda_{n^r,j}&=&  -\frac{R^2E_{n^r}}{2}  + \frac{(2 n^r+1) R(Z_1+Z_2)}{2 n^r +2} -q_j-\frac{1}{4}\nonumber\\F_{n^r,j}(\xi)&=& \sqrt{\xi-1} \, e^{- \frac{R(Z_1+Z_2)}{2n^r+2}\xi} u_{n^r,j}(\xi)\nonumber
\end{eqnarray}

In the cases c) and d) the energies do not correspond to the planar hydrogenoid atoms, because the even character of the denominator, $2n^r+2$. Recall that the spectrum of the planar hydrogen problem\cite{GGTCM} is: $E=\frac{-2}{(2n+1)^2}$. Nevertheless, there exist
interesting elementary solutions of Types c) and d) which are reminiscent of the lemniscatic orbits of the classical problem and we shall
refer to as \lq\lq quasi\rq\rq hydrogenoid.

\section{Elementary solutions of Whittaker-Hill equation}

An equivalent analysis can be performed for the angular equation (\ref{angular2d}) simply changing the relative sign of the charge $Z_1$ and the range of variation for the angular coordinate. Thus, the condition: $\alpha= - n^a \epsilon$, $n^a \in {\mathbb N}$, leads to the following structure of solutions:

\textbf{Type a}. $\delta=\gamma=\frac{1}{2}$.
\begin{eqnarray}
 E_{n^a}&=&-\frac{2 (-Z_1+Z_2)^2}{(2 n^a +1)^2}\nonumber\\ \lambda_{n^a,j}&=& -\frac{R^2E_{n^a}}{2} + \frac{2 n^a R(-Z_1+Z_2)}{2 n^a +1} -q_j\nonumber\\ G_{n^a,j}(\eta)&=& e^{- \frac{R(-Z_1+Z_2)}{2n^a+1}\eta} u_{n^a,j}(\eta)\nonumber
\end{eqnarray}

\textbf{Type b}. $\delta=\gamma=\frac{3}{2}$.
\begin{eqnarray}
 E_{n^a}&=&-\frac{2 (-Z_1+Z_2)^2}{(2 n^a +3)^2}\nonumber\\ \lambda_{n^a,j}&=& -\frac{R^2E_{n^a}}{2} + \frac{2 n^r R(-Z_1+Z_2)}{2 n^a +3} -q_j-1\nonumber\\ G_{n^a,j}(\eta)&=& \sqrt{1-\eta^2}\,  e^{- \frac{R(-Z_1+Z_2)}{2n^a+3}\eta} u_{n^a,j}(\eta)\nonumber
\end{eqnarray}

\textbf{Type c}. $\delta=\frac{1}{2}$, $\gamma=\frac{3}{2}$.
\begin{eqnarray}
 E_{n^a}&=&-\frac{2 (-Z_1+Z_2)^2}{(2 n^a +2)^2}\nonumber\\ \lambda_{n^a,j}&=& -\frac{R^2E_{n^a}}{2} + \frac{(2 n^a-1) R(-Z_1+Z_2)}{2 n^a +2} -q_j-\frac{1}{4}\nonumber\\G_{n^a,j}(\eta)&=& \sqrt{1+\eta} \, e^{- \frac{R(-Z_1+Z_2)}{2n^r+2}\eta} u_{n^a,j}(\eta)\nonumber
\end{eqnarray}

\textbf{Type d}. $\delta=\frac{3}{2}$, $\gamma=\frac{1}{2}$.
\begin{eqnarray}
 E_{n^a}&=&-\frac{2 (-Z_1+Z_2)^2}{(2 n^a +2)^2}\nonumber\\ \lambda_{n^a,j}&=&  -\frac{R^2E_{n^a}}{2}  + \frac{(2 n^a+1) R(-Z_1+Z_2)}{2 n^a +2} -q_j-\frac{1}{4}\nonumber\\G_{n^a,j}(\eta)&=& \sqrt{1-\eta} \, e^{- \frac{R(-Z_1+Z_2)}{2n^a+2}\eta} u_{n^a,j}(\eta)\nonumber
\end{eqnarray}

\section{Elementary solutions of the Schr\"odinger equation}

The simultaneous existence of elementary solutions in both radial and angular equations, for fixed values of $Z_1$ and $Z_2$, is obviously possible only if the parameters $E$ and $\lambda$ determined in the resolution process are the same for both equations.

Thus, there exist elementary solutions of the Schr$\ddot{\rm o}$dinger equation only for those values of $n_1\in {\mathbb N}$ and $n_2\in {\mathbb N}$ that solve the diophantine equation:

\begin{equation}
\frac{(Z_1+Z_2)^2}{n_1^2}=\frac{(-Z_1+Z_2)^2}{n_2^2}\, , \label{diophantine}
\end{equation}
where $n_1$ can be equal to $2n^r+1$, $2n^r+3$ or $2n^r+2$, whereas $n_2$ is: $2n^a+1$, $2n^a+3$ or $2n^a+2$. Moreover, the value of the separation constant $\lambda$ in (\ref{radial2d}) and (\ref{angular2d}) must be the same and because of the dependence of $\lambda$ on $R$, the equality will only be satisfied for certain values of the internuclear distance.

As illustrative examples, we present several elementary solutions for two pairs of charges: $Z_1=5$, $Z_2=1$, and $Z_1=2$, $Z_2=1$.

\noindent {\it Charges $Z_1=5$, $Z_2=1$}. The elementary eigenfunction of minimum energy corresponds to the $n_1=3$ and $n_2=2$ solution of (\ref{diophantine}): $E=-8$. This solution appears considering Type b) in Razavy equation with $n^r=0$ and Type d) in Whittaker-Hill equation with $n^a=0$. The compatible value of $\lambda$ is $-7/16$ that is obtained for $R=3/8$. Thus, we have the eigenfunction:

\begin{equation}
\Psi(\xi,\eta) = \sqrt{\xi^2-1}\sqrt{1-\eta} e^{-\frac{3}{4}(\xi-\eta)}\label{example1}
\end{equation}

\begin{figure}
\begin{center}
\includegraphics[height=4cm]{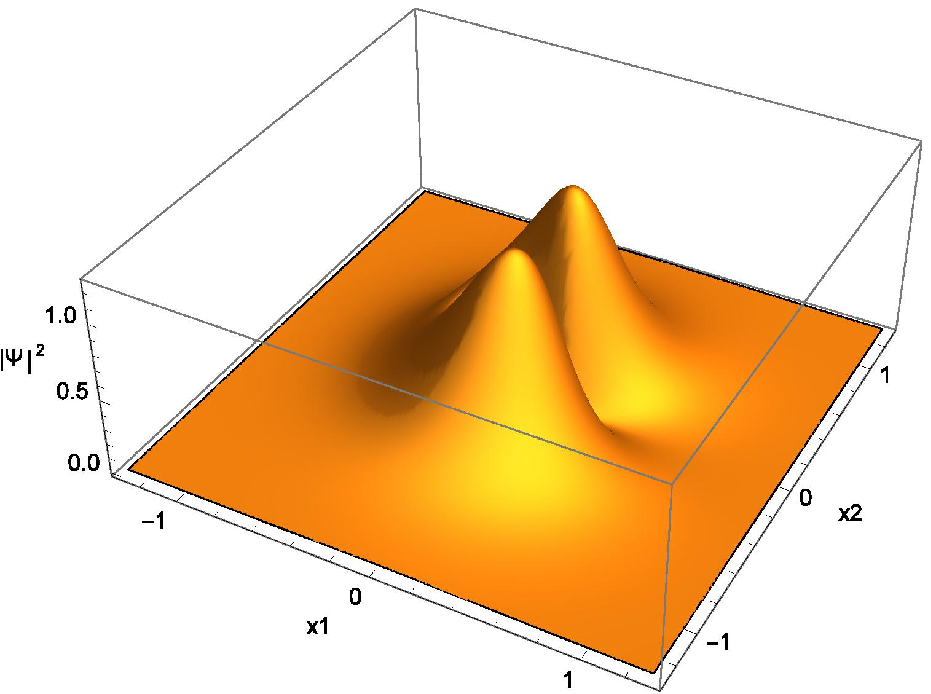}
\includegraphics[height=4cm]{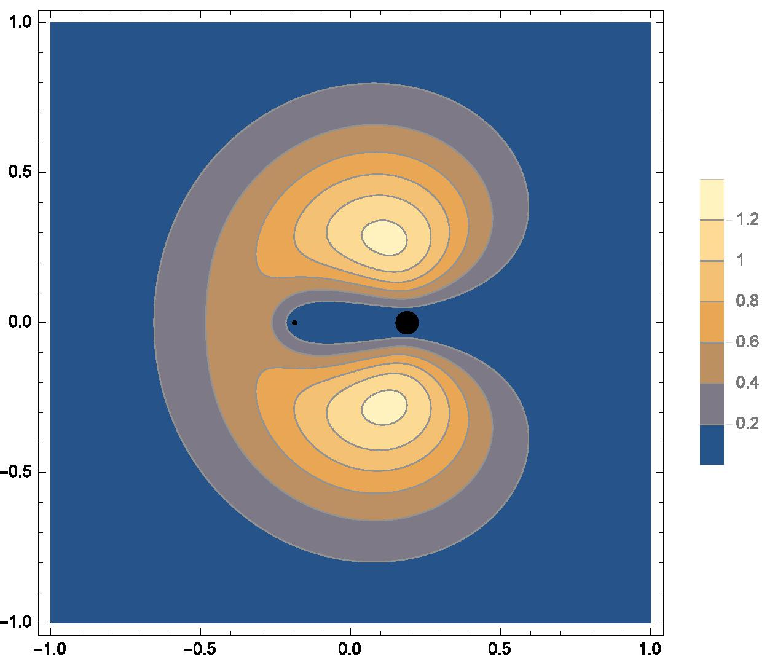}
\caption{\small Graphical representation of the probability density $\rho(x_1,x_2)=\left| \Psi \right|^2$ for the normalized eigenfunction (\ref{example1}), and several level curves of $\rho(x_1,x_2)$.}
\label{fig.1}
\end{center}
\end{figure}

The next energy level where elementary solutions are found is $E=-2$. These solutions are obtained for $n_1=6$ and $n_2=4$ in (\ref{diophantine}), in four different combination of Types c) and d):

\begin{itemize}

\item Type d) in both equations. The values of the constants are:  $\lambda=-583/256$ and $R=3/16$ and the wave function reads:
\begin{eqnarray}
\Psi(\xi,\eta) &=& \sqrt{\xi -1} \sqrt{1-\eta } \left(\eta +\frac{1}{3}\right) \nonumber\\ && \times \left(\xi ^2-10
   \xi -7\right) e^{-\frac{3}{16}(\xi -\eta )}\label{example2}
\end{eqnarray}

\item Type d) in Razavy equation and Type c) in Whittaker-Hill equation. $\lambda=-2.247$, $R=0.435$
\begin{eqnarray}
\Psi(\xi,\eta)&=&\sqrt{\xi -1}\sqrt{1+\eta}(\eta -0.713)\nonumber\\ && \times \left(\xi ^2-3.889 \xi -3.538 \right) e^{- 0.435 (\xi-\eta)}\nonumber
\end{eqnarray}

\item Type c) and d) respectively. $\lambda=-3.111$,  $R=1.643$
\begin{eqnarray}
\Psi(\xi,\eta)&=& \sqrt{\xi +1} \sqrt{1-\eta } (\eta -0.583) \nonumber\\ && \times \left(\xi^2 - 0.634 \xi
   -0.881 \right) e^{ -1.643(\xi -\eta) }\nonumber
   \end{eqnarray}

\item Type c) and d) again. $\lambda=-0.587$,  $R=0.292$
\begin{eqnarray}
\Psi(\xi,\eta)&=& \sqrt{\xi +1} \sqrt{1-\eta } (\eta +3.203) \nonumber\\ & &\times
 \left(\xi ^2-10.055 \xi  + 16.819 \right) e^{-0.292 (\xi -\eta) }\nonumber
 \end{eqnarray}

\end{itemize}

\begin{figure}
\begin{center}
\includegraphics[height=4cm]{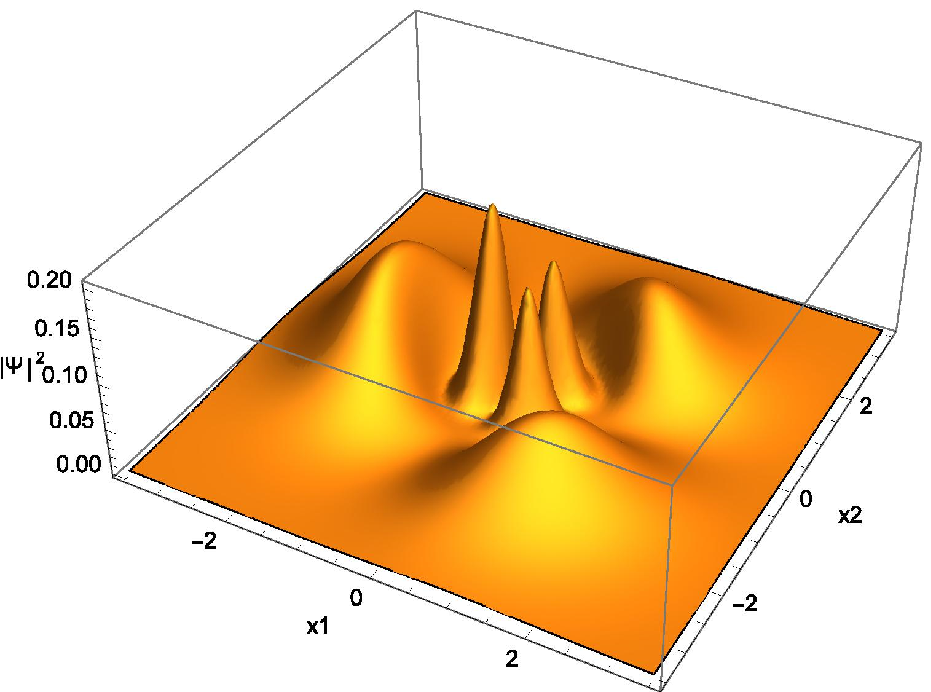}
\includegraphics[height=4cm]{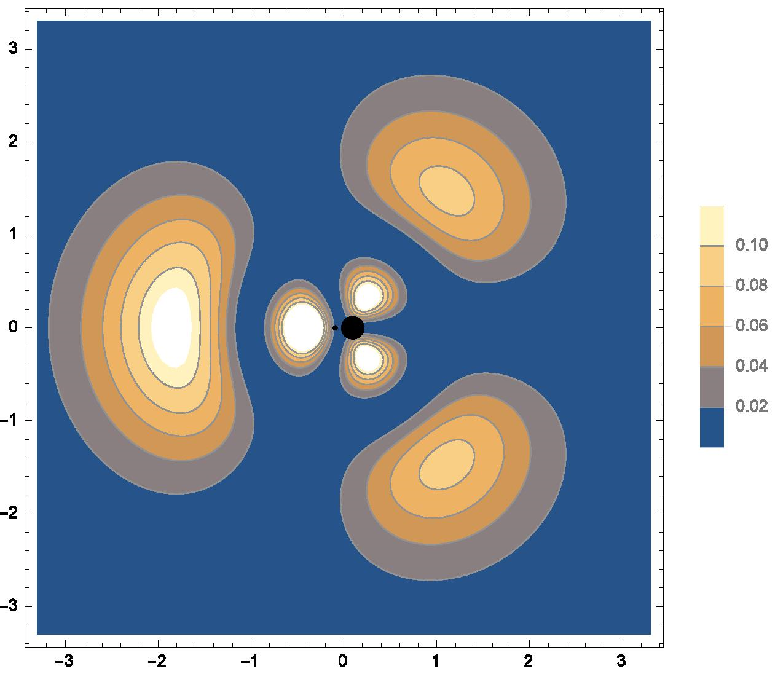}
\caption{\small Graphical representation of $\rho(x_1,x_2)$ for the normalized eigenfunction (\ref{example2}) and several level curves of $\rho(x_1,x_2)$.}
\label{fig.2}
\end{center}
\end{figure}

\noindent {\it Charges $Z_1=2$, $Z_2=1$}. In this case we present the three elementary eigenfunctions that appears for: $E=-1/2$. The corresponding solution of (\ref{diophantine}) is $n_1=6$ and $n_2=2$.

\begin{itemize}

\item $\lambda= \frac{1}{36}(59-4 \sqrt{19})$, $R=\frac{2}{3}(1+\sqrt{19})$
 \begin{eqnarray}
 \Psi(\xi,\eta)&=& \sqrt{\xi -1} \sqrt{1-\eta } e^{-\frac{1}{3}
   \left(1+\sqrt{19}\right) (\xi -\eta)}\nonumber\\ &\times &\left(\xi ^2+\frac{1}{2} \left(3-\sqrt{19}\right) \xi +\frac{1}{4} \left(3-2 \sqrt{19}\right) \right)\nonumber
\end{eqnarray}

\item $\lambda=-\frac{31}{64}$, $R=\frac{3}{4}$

\begin{equation}
\Psi(\xi,\eta)=  \sqrt{\xi +1} \sqrt{1-\eta } (\xi^2 -8\xi +11) e^{-\frac{3}{8}(\xi -\eta )}\nonumber
\end{equation}

\item $\lambda=\frac{1}{36}(59+4 \sqrt{19})$, $R=\frac{2}{3}(-1+\sqrt{19})$

\begin{eqnarray}
\Psi(\xi,\eta)&=&\sqrt{\xi +1} \sqrt{1+\eta}   e^{-\frac{1}{3} (-1+\sqrt{19})(\xi- \eta)}\nonumber\\ & \times& \left(\xi ^2-\frac{1}{2} \left(3+\sqrt{19}\right) \xi +\frac{1}{4} \left( 3+2\sqrt{19}\right)\right)\nonumber
\end{eqnarray}
\end{itemize}
In order to facilitate the presentation we have written some of the numbers rounded to three significant decimal digits, although all these elementary eigenfunctions have been obtained using exact analytic expressions. There exist also elementary solutions containing polynomials of higher orders, that have to be calculated using numerical approximations because the determination of the roots of ${\cal P}_{n+1}(q)$ in (\ref{recurrence}) with $n\geq 4$ is involved.

\section{Two-equal centers}

It is very interesting to analyze the symmetric case, i.e., when $Z_1=Z_2=Z$. The radial equation (\ref{radial2d}) is still a Razavy equation, but the angular one (\ref{angular2d}) reduces to the algebraic version of Mathieu equation:

\begin{equation}
\label{MathieuA}
(1-\eta^2)\frac{d^2G(\eta)}{d\eta^2}- \eta \frac{dG(\eta)}{d\eta} - \left( \frac{ER^2}{2} \eta^2+\lambda\right) G(\eta) = 0
\end{equation}
The standard form of Mathieu equation \cite{Whittaker, Arscott} is obtained by the change of variable: $\eta\, =\, \cos \nu$, where $\nu$
varies in the interval $\nu\in [-\pi,\pi]$ in such a way that the $x_2>0$ half-plane is described in the range $\nu\in(0,\pi)$ whereas $\nu\in (-\pi,0)$ corresponds to the $x_2<0$ half-plane, i.e., the original change of coordinates:

\begin{equation}
x_1 =  \frac{R}{2}  \xi \eta\  ,\quad  x_2= \pm  \frac{R}{2}  \sqrt{\xi^2-1} \sqrt{1-\eta^2}\nonumber
\end{equation}
is redefined in a one-to-one version:

\begin{equation}
x_1 =  \frac{R}{2}  \xi\, \cos \nu\  ,\quad  x_2=  \frac{R}{2}  \sqrt{\xi^2-1}  \sin \nu \nonumber
\end{equation}
Thus, eq. (\ref{MathieuA}) reduces to standard form:

\begin{equation}
\frac{d^2G(\nu)}{d\nu^2}+  \left( a - 2p\, \cos 2\nu \right) G(\nu) = 0\label{Mathieu1}
\end{equation}
where $a =  -\lambda - \frac{ER^2}{8}$ and $p=\frac{ER^2}{8}$.

Mathieu equation (\ref{Mathieu1}) does not admit polynomial solutions \cite{Arscott}, this fact can be easily checked because the quasi-exact solvability condition for CHEq is not compatible with the form of Mathieu equation (\ref{MathieuA}). Complete elementary solutions, products of two polynomials, for the planar two center problem in the symmetric case do not exist.

Nevertheless, it is possible to consider elementary solutions only for Razavy equation and to determine the solutions of Mathieu equation corresponding to the values of $E_{n^r}$ and $\lambda_{n^r,j}$, (\ref{energyRaz},\ref{lambdaj}). Each elementary Razavy solution is associated with a Mathieu equation (\ref{Mathieu1}) with the parameters:

\begin{equation}
a_{n^r,j} = - \lambda_{n^r,j} - \frac{E_{n^r} R^2}{4} \  , \quad p_{n^r} = \frac{E_{n^r}R^2}{8}\nonumber
\end{equation}

There exist solutions of Mathieu equation (\ref{Mathieu1}) with different periodicity or quasi-periodicity properties, and they are applicable to a physical problem depending on its special characteristics, see for instance \cite{Condon}. In the two-center problem, the physically admissible eigenfunctions of the Schr\"odinger equation must be univalued and thus it is necessary to consider only the $2\pi$-periodic solutions of (\ref{Mathieu1}), the proper Mathieu functions: $Ce_n(p,z)$ and $Se_n(p,z)$ \cite{Whittaker, Arscott}.
The Mathieu cosine $Ce_n(p,z)$, $n=0,1,\dots$, and sine $Se_n(p,z)$, $n=1,2,\dots$, are solutions of (\ref{Mathieu1}) only if the parameter $a$ takes the Mathieu characteristic values $a_n(p)$ and $b_n(p)$ respectively.

\begin{figure}
\begin{center}
\includegraphics[height=4cm]{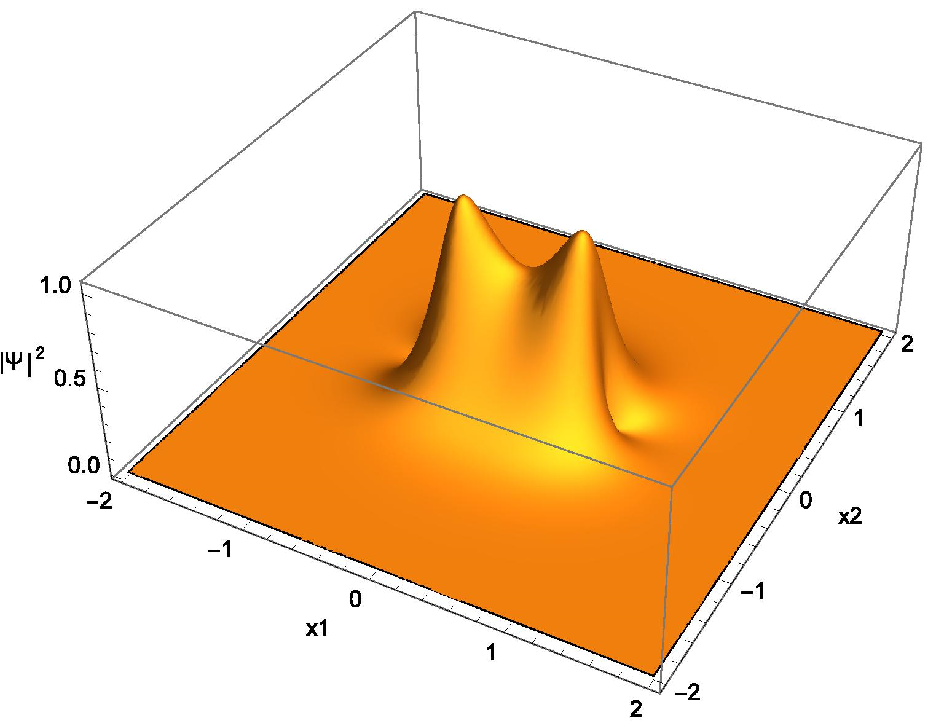}
\includegraphics[height=4cm]{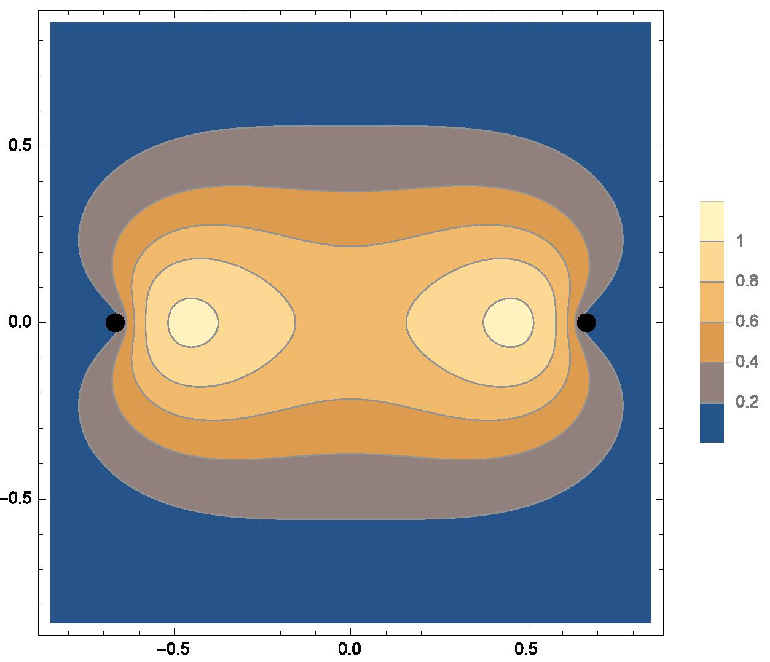}
\caption{\small Graphical representation of $\rho(x_1,x_2)$ for (\ref{mat1}) and several level curves of $\rho(x_1,x_2)$.}
\end{center}
\end{figure}

An elementary Razavy solution determined by $E_{n^r}$ and $\lambda_{n^r,j}$ is compatible only with values of $R$ such that the Mathieu parameter $a_{n^r,j}$ coincides with a characteristic value of the Mathieu cosine or sine functions, i.e.

\begin{equation}
-\lambda_{n^r,j}-\frac{E_{n^r}R^2}{4} = a_n\left( \frac{E_{n^r}R^2}{8}\right)\label{matha}
\end{equation}
for some $n=0,1,2,\dots$, or:

\begin{equation}
-\lambda_{n^r,j}-\frac{E_{n^r}R^2}{4} = b_n\left( \frac{E_{n^r}R^2}{8}\right)\label{mathb}
\end{equation}
for some $n=1,2,\dots$.

As examples, several eigenfunctions of this type are presented:

\noindent {\it Charges $Z_1=Z_2=3$}. We illustrate the process with the simplest example: Choosing the energy $E=-8$, obtained for $n^r=1$ in the Type a) of Razavy equation, there are two allowable values for the separation constant: $\lambda_{1,1}$ and $\lambda_{12}$.

The associated Mathieu equations have parameters: $a_{1,1}$ and $a_{1,2}$ respectively, and $p_1=-R^2$. There exists only one possibility for $R$ that leads $a_{1,1}$ and/or $a_{1,2}$ to be a characteristic value of the Mathieu equation: if $R=1.335$, then $a_{1,2}= b_1(p_1)=2.298$, $\lambda_{1,2}=1.268$. For these values, the trascendent equation (\ref{mathb}) is satisfied. Thus the eigenfunction is:

\begin{equation}
\Psi(\xi,\nu)= e^{-2.671 \xi } (\xi +0.911) Se_1(-1.783,\nu)\label{mat1}
\end{equation}

Other examples are:

\begin{itemize}
\item $E=-18$, $\lambda=-0.264$, $R=0.329$, $a_1(p_0)=0.750$.

\begin{equation}
\Psi(\xi,\nu)= e^{-0.986 \xi } \sqrt{\xi +1} Ce_1(-0.243,\nu )\label{mat2}
\end{equation}

\item $E=-\frac{9}{2}$, $\lambda=-3.133$, $R=0.870$, $b_2(p_1)=3.985$.

\begin{equation}
\Psi(\xi,\nu)= e^{-1.305 \xi } \sqrt{\xi +1} (\xi +0.491)
   Se_2(-0.426,\nu)\nonumber
\end{equation}

\item $E=-\frac{72}{25}$, $\lambda=6.412$, $R=4.491$, $a_2(p_2)=8.111$.

\begin{eqnarray}
\Psi(\xi,\nu)&=& e^{-5.389 \xi } \left(\xi ^2+1.729 \xi +0.735\right)\nonumber\\
&&\times  Ce_2(-7.262,\nu )\nonumber
\end{eqnarray}
\end{itemize}
For aesthetic reasons, the characteristic values and the other quantities involved in (\ref{mat1},\ref{mat2}) and rest of equations, are presented using only three significant decimal digits, although obviously an arbitrary precision can be obtained.

It is possible to find a plethora of solutions of this mixed type with specific values of $R$ physically meaningful in the atomic/molecular range.

\begin{figure}
\begin{center}
\includegraphics[height=4cm]{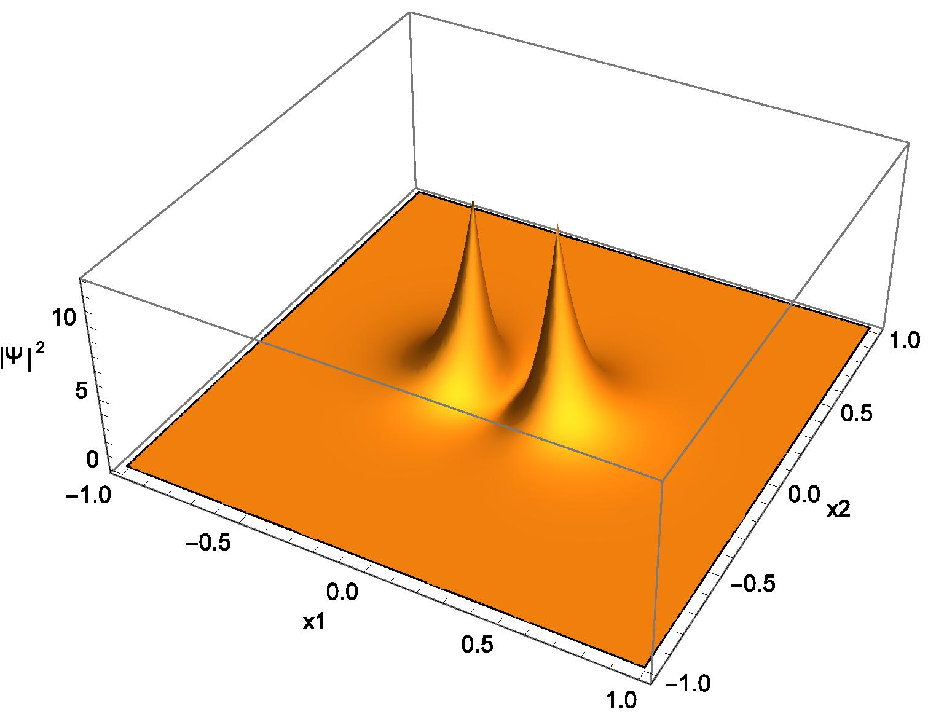}
\includegraphics[height=4cm]{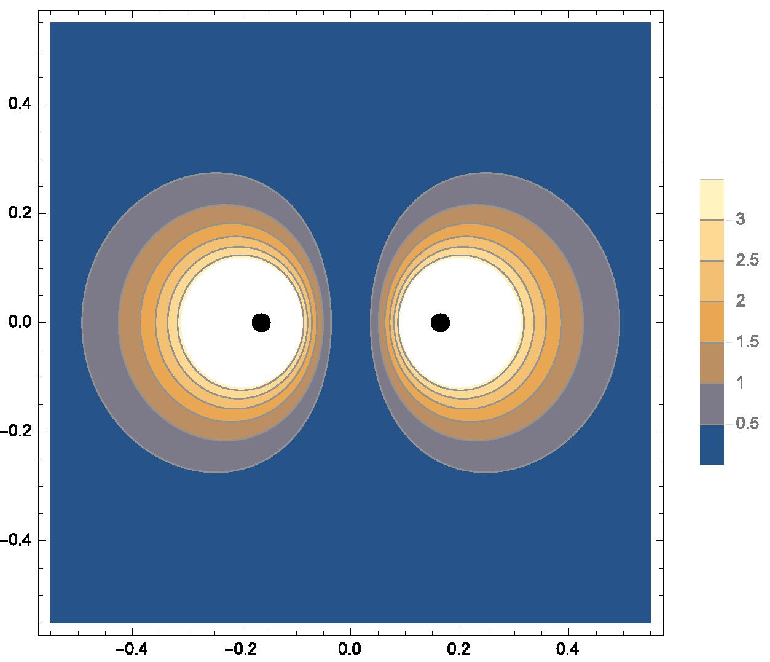}
\caption{\small Graphical representation of $\rho(x_1,x_2)$ for (\ref{mat2}) and several level curves of $\rho(x_1,x_2)$.}
\label{fig.4}
\end{center}
   \end{figure}

\section*{Acknowledgments}

The authors thank the Spanish Ministerio de Econom\'{\i}a y Competitividad for financial support under grant MTM2014-57129-C2-1-P.

\end{document}